\documentclass[aps,pre,reprint,floatfix,superscriptaddress]{revtex4-1}

\usepackage{graphicx}
\usepackage{amsmath,amssymb,amsfonts}
\usepackage{hyperref}
\usepackage{url}
\usepackage{numprint}

\begin{document}

\title{Breaking universality in random sequential adsorption on a square lattice with long-range correlated defects}

   \author{Sumanta Kundu}
   \email{sumanta@spin.ess.sci.osaka-u.ac.jp}
   \affiliation{Department of Earth and Space Science, Osaka University, 560-0043 Osaka, Japan}
   \author{Dipanjan Mandal}
   \email{dipanjan.mandal@warwick.ac.uk}
   \affiliation{ Department of Physics, University of Warwick, Coventry CV4 7AL, United Kingdom}

\date{\today} 
\begin{abstract}

   Jamming and percolation transitions in the standard random sequential adsorption of 
   particles on regular lattices are characterized by a universal set of critical 
   exponents. The universality class is preserved even in the presence of randomly 
   distributed defective sites that are forbidden for particle deposition. However, 
   using large-scale Monte Carlo simulations by depositing dimers on the square 
   lattice and employing finite-size scaling, we provide evidence that the system 
   does not exhibit such well-known universal features when the defects have spatial 
   long-range (power-law) correlations. The critical exponents $\nu_j$ and $\nu$ 
   associated with the jamming and percolation transitions, respectively, are found 
   to be non-universal for strong spatial correlations and approach systematically their 
   own universal values as the correlation strength is decreased. More crucially, 
   we have found a difference in the values of the percolation correlation length 
   exponent $\nu$ for a small but finite density of defects with strong spatial 
   correlations. Furthermore, for a fixed defect density, it is found that the 
   percolation threshold of the system, at which the largest cluster of absorbed 
   dimers first establishes the global connectivity, gets reduced with increasing 
   the strength of the spatial correlation. 
\end{abstract}

\maketitle

\section{Introduction}

   The study of adsorption of particles onto solid surfaces is a subject of great
   interest in different disciplines of science and technology
   \cite{Evans1993,Feder1980,Torquato2010,Kumacheva2002} due to its 
   relevance in diverse applications, including protein adsorption~\cite{Hlady1996}, 
   ion implantation in semiconductor~\cite{Roman1983}, and thin film deposition 
   technologies for surface coatings and encapsulations~\cite{Yu2016}. In the
   simplest case of adsorption leading to monolayer formation, such as the binding 
   of protein molecules on glass or metals~\cite{Feder1980}, one considers that the 
   process of adsorption takes place irreversibly and the particles have no mobility. 
   Consequently, they remain at their position of adsorption forever. However, many 
   complex dynamical phenomena, such as diffusion, desorption, and thermal expansion 
   of particles are often found to be associated with the process of adsorption 
   occurring in the real-world systems~\cite{Privman1994,Ramsden1992,Joshi2016}. It 
   has been observed that such underlying mechanisms crucially affect the morphology 
   of the growing monolayer formations. Apart from that, properties of the surface, 
   for example, the surface roughness or the interfacial interaction play a significant 
   role in the kinetics of adsorption~\cite{Napolitano2020}. However, to our knowledge, 
   the latter aspects have not been studied in great detail using theoretical models.
   
   The theoretical study of monolayer formation in the limit of irreversible adsorption 
   has been carried out quite intensively over the last several decades through the
   stochastic models of random sequential adsorption (RSA)
   \cite{Privman2000,Cadilhe2007,Evans1993}. In the standard RSA, particles are absorbed 
   sequentially and irreversibly at random positions onto an initially empty substrate,
   subject to a constraint that they only interact through excluded volume interaction. 
   The kinetics of adsorption terminates when a jamming state is reached where no more 
   vacant space is available to accommodate a single particle. The surface coverage 
   $p$, defined as the volume fraction of the surface occupied by the adsorbed particles, 
   attains a non-trivial value $p_j$ at the jamming limit. The exact value of $p_j$ is 
   known only for one-dimensional systems in both continuum and lattice spaces
   \cite{Flory1939,Renyi1958}.
 
   Another important aspect which has been studied using the RSA model is the phenomenon 
   of percolation of polyatomic species 
   \cite{Kondrat2001,Cornette2003,Tarasevich2012,Gimenez2015}. A group of adsorbed 
   particles, occupying more than one lattice site, forms clusters through their 
   neighboring connections. The percolation transition occurs when such a cluster 
   connects two opposite boundaries of the system through a spanning path at a 
   critical value of the surface coverage $p = p_c$, known as the percolation 
   threshold. Therefore the global connectivity exists in the system only in the 
   percolating phase of $p > p_c$. The system exhibits the generic scale-invariant 
   features of a continuous phase transition right at $p_c$~\cite{Stauffer2018}.
   
   Furthermore, the role played by the shape and size of the depositing particles
   on the morphology of the growing structure has been studied
   \cite{Becklehimer1992,Kondrat2001,Lebovka2011,Tarasevich2012,ziff2016}. Different 
   mechanisms of adsorption have also been introduced to explain various experimental 
   observations comprehensively~\cite{Guo1994,Joshi2016,Kundu2018,Furlan2020}. It has 
   been revealed that the jamming density $p_j$ and the percolation threshold $p_c$ depend 
   non-trivially on all these factors. However, interestingly, the critical behavior 
   of the system associated with the two transition points is found to be universal, 
   meaning that it is characterized by a universal set of critical exponents that 
   is independent of all these microscopic details. In all these cases, the percolation 
   transition belongs to the ordinary percolation universality class \cite{Cornette2003,Tarasevich2012,Lebovka2011,Gimenez2015,Kundu2018,Furlan2020}. 
   Similarly, the jamming transition is characterized by the universal exponent $\nu_j$ 
   relating to the size scaling of the width $\Delta$ of the transition zone, which 
   scales with linear size $L$ of the substrate in a spatial dimension $d$ as $\Delta 
   \sim L^{-1/\nu_j}$, with $\nu_j = 2/d$. The robustness of this universal scaling law 
   has been examined on the Euclidean and fractal lattice geometries~\cite{Pasinetti2019,Ramirez2019}.
   
   Although the effect of particle properties on the kinetics of adsorption have been 
   extensively studied in the past, a theoretical investigation on the role of surface 
   properties has remained almost unexplored. In this context, some previous studies 
   have considered that the surface on which the adsorption is taking place is not 
   ideal. It contains defects or impurities at random places 
   \cite{Ramirez2019,Cornette2006,Kondrat2006,Centres2015,Tarasevich2015,Palacios2020},
   indicating that the binding strength of particles at these locations is so negligible that 
   they can not be attached there. Except for these places the adsorption is possible 
   if the vacant space is large enough to accommodate a particle. Even in this case, 
   the universal behavior of the RSA model is preserved. However, in many realistic 
   situations, a surface shows spatially correlated properties~\cite{Medina1989,Hewett1990,Lauritsen1993} 
   and thus, the existence of spatial correlations among the defects is a more 
   natural consideration than the randomly distributed defects. Surprisingly, 
   this aspect has not been considered yet.
   
   In this paper, we provide a detailed study on the jamming and percolation 
   properties of the RSA model in the presence of spatially long-range (power-law) 
   correlated defects. Our main interest is to see whether this spatial correlation 
   affects the critical behavior of the transitions. We found that in the regime of 
   non-vanishing spatial correlation among the defects, the model does not exhibit 
   the well-known universal features of the RSA. The critical exponents associated with 
   the jamming and percolation transitions are observed to vary systematically with 
   the strength of the spatial correlation.
   


\section{Model}
 
   We consider the RSA model on a two dimensional square lattice of size $L \times 
   L$ with periodic boundary conditions along both directions. Particles in the form 
   of $k$-mers, occupying $k$ consecutive lattice sites along horizontal or vertical 
   direction, are adsorbed one by one following the rules of the standard RSA onto 
   the lattice consisting of defects which are spatially correlated. Specifically, 
   the defective sites are located in such a 
   manner that the correlation function between a pair of defected sites decays in 
   the form of a power-law of the spacial distance between them. By imposing this 
   prerequisite, sites of an empty lattice are occupied with probability $q$ and 
   they are kept vacant with probability $1-q$ (the detailed method for generating 
   such a correlated landscape is described later). The initial configuration of these 
   sets of occupied sites act as defects and the adsorption of particles on these 
   locations is completely forbidden. The remaining $1-q$ fraction of sites acts as the 
   site for possible adsorption. By selecting an orientation, either horizontal or 
   vertical with equal probability, one end of the particle is placed at a randomly 
   selected position and absorbed irreversibly provided that there exists at least 
   $k$ consecutive vacant sites along the chosen direction from the selected site. 
   The surface coverage after adsorbing $n$ particles is thus given by $p = nk/L^2$. 
   The adsorption process continues in this way until a jamming state is reached. 
   The corresponding surface coverage $p=p_j$ is referred to as the jamming density.
   
   During the process of adsorption, the depositing particles interact among the 
   previously adsorbed particles as well as with the defects via excluded volume 
   interaction. Consequently, they experience a ``screening effect'' and try to align 
   parallel to each other and form domains whose typical sizes are of the order of 
   the size of the particle. In addition, due to this interaction, a vacant region 
   that is smaller than the size of the particle can not be occupied. As a result, 
   the jamming density $p_j$ can not attain the close packing density, i.e., $p_j
   < 1-q$. 
   
   As an important step, besides the defect density parameter $q$, the strength of 
   the correlation among the defects is also a tunable parameter in our model and 
   it is characterized by an exponent $\gamma$ associated with the power-law decay 
   of the correlation function (see Eq.~(\ref{spacial_correlation})). For any 
   arbitrary value of $0 < q < 1$, the strength of the correlation decreases with 
   increasing the value of $\gamma$ and in the limit of $\gamma \to \infty$ the 
   scenario of uncorrelated defects is obtained. Therefore, by varying the parameters 
   $q$ and $\gamma$, the model is capable of capturing the behavior of a wide range 
   of systems: from a system with correlated defects, uncorrelated defects to a pure 
   (defect-free) system. In this paper, we report our simulation results for dimers 
   $(k = 2)$.
   
   
%
    \begin{figure}[b]
    \centering
    \includegraphics[width=0.98\linewidth]{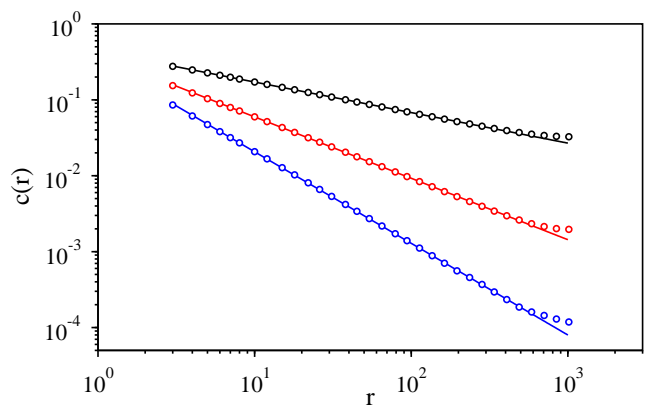}
    \caption{Log-log plot of the height-height correlation function $c({\bf r})$ 
    against the spatial distance $|{\bf r}|$ on a two-dimensional surface with 
    $L = 2^{11}$ using $\gamma$ = 0.4 (black), 0.8 (red), and 1.2 (blue) (open 
    circles). By averaging over $10^4$ independent configurations, the slope of 
    the fitted straight lines have been estimated as 0.408(5), 0.804(5), and 
    1.206(5), respectively (arranged from top to bottom).}
    \label{corrfunc}
    \end{figure}
%
\subsection*{Generating long-range correlated defects}

   To obtain a substrate that possesses defects with spatial long-range correlations, we utilize
   the idea of viewing the substrate as a landscape of random heights $\{h({\bf x})\}$ with 
   desired height-height correlation~\cite{Schrenk2013}, where $h({\bf x})$ represents the
   height associated with the lattice site positioned at ${\bf x}$. Accordingly, we follow 
   the scheme described in Ref.\ \cite{Makse1996}, which is based on the Fourier filtering 
   method~\cite{Makse1996,Prakash1992,Zierenberg2017}. The Wiener-Khintchine theorem is the 
   basis of this method, which relates the autocorrelation function of a stationary time 
   series to the Fourier transform of its power spectrum. The power spectral density 
   in this case has a power-law form and it is calculated using the two-point correlation 
   function $c({\bf r}) = (1+|{\bf r}|^2)^{-\gamma/2}$ imposing periodic boundary conditions 
   in two dimensions. Therefore, $c({\bf r})$ decays at large distance $|{\bf r}|$ as
   \begin{equation}
   c({\bf r}) = \langle h({\bf x})h({\bf r} + {\bf x}) \rangle \sim |{\bf r}|^{-\gamma}
   \label{spacial_correlation}
   \end{equation}
   where, $\gamma$ denotes the strength of the correlation. The steps (i)--(iii) in Ref.\ 
   \cite{Makse1996} are then executed to generate the correlated random numbers $\{h({\bf x})\}$.
   
   In our simulation we used Gaussian distributed uncorrelated random numbers with 
   zero mean and unit variance to generate power-law correlated Gaussian distributed 
   random numbers. To verify whether the obtained random numbers posses the desired 
   correlations or not, the configuration averaged value of $c({\bf r})$ is plotted 
   with $|{\bf r}|$ on a double logarithmic scale in Fig.\ \ref{corrfunc} for three 
   different values of $\gamma$. Clearly, the measured slopes of the best fitted 
   straight lines are consistent with the desired values of $\gamma$. 
   
   Finally, by following the idea of ranked surface~\cite{Schrenk2012}, the sites are 
   occupied one by one according to the ascending order of their height values until the 
   density of occupied sites representing the defects reaches a prefixed value $q$.
    
\begin{figure}[t]
\centering
\begin{tabular}{cc}
\includegraphics[width=0.47\linewidth]{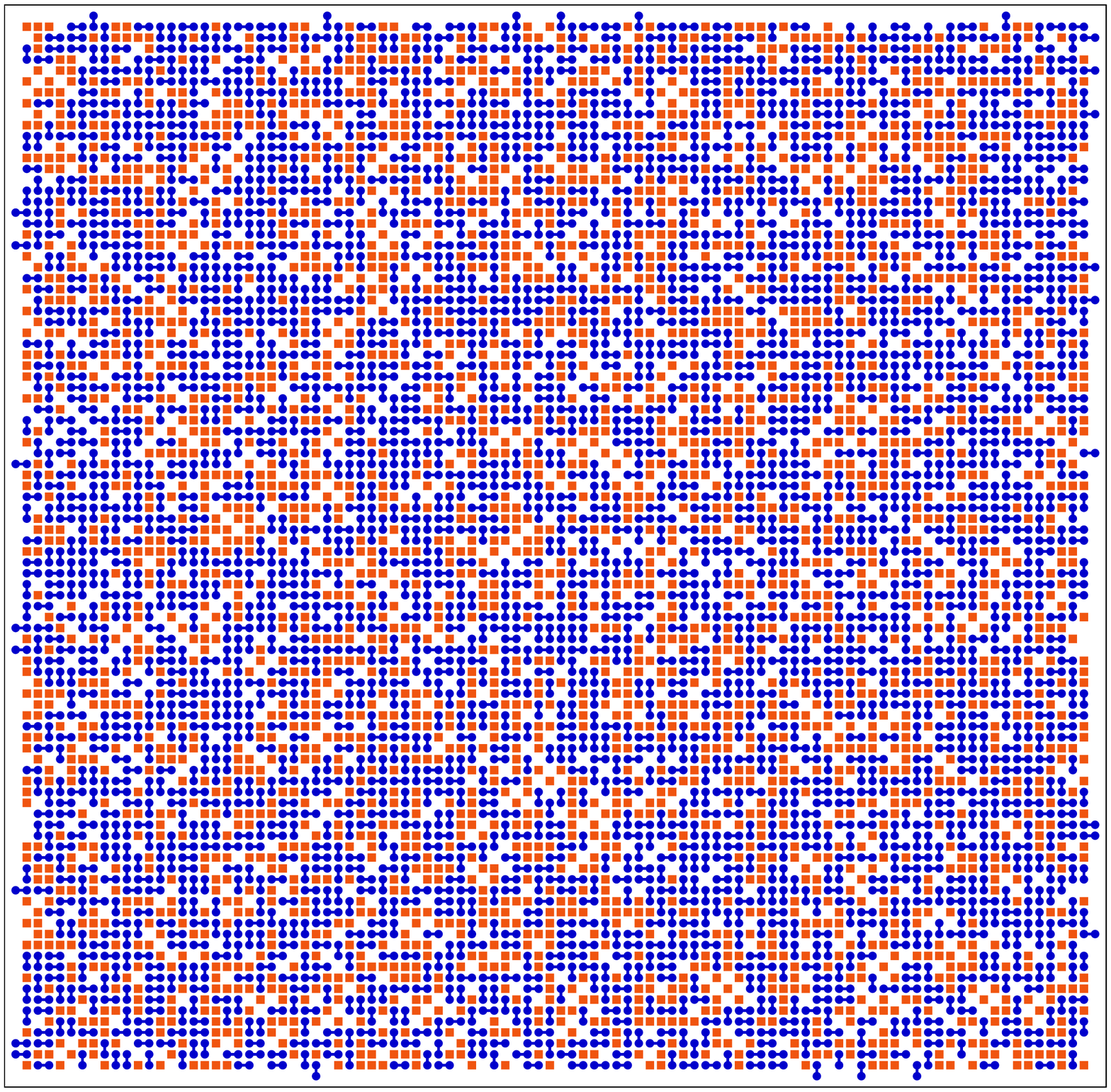} & 
\includegraphics[width=0.47\linewidth]{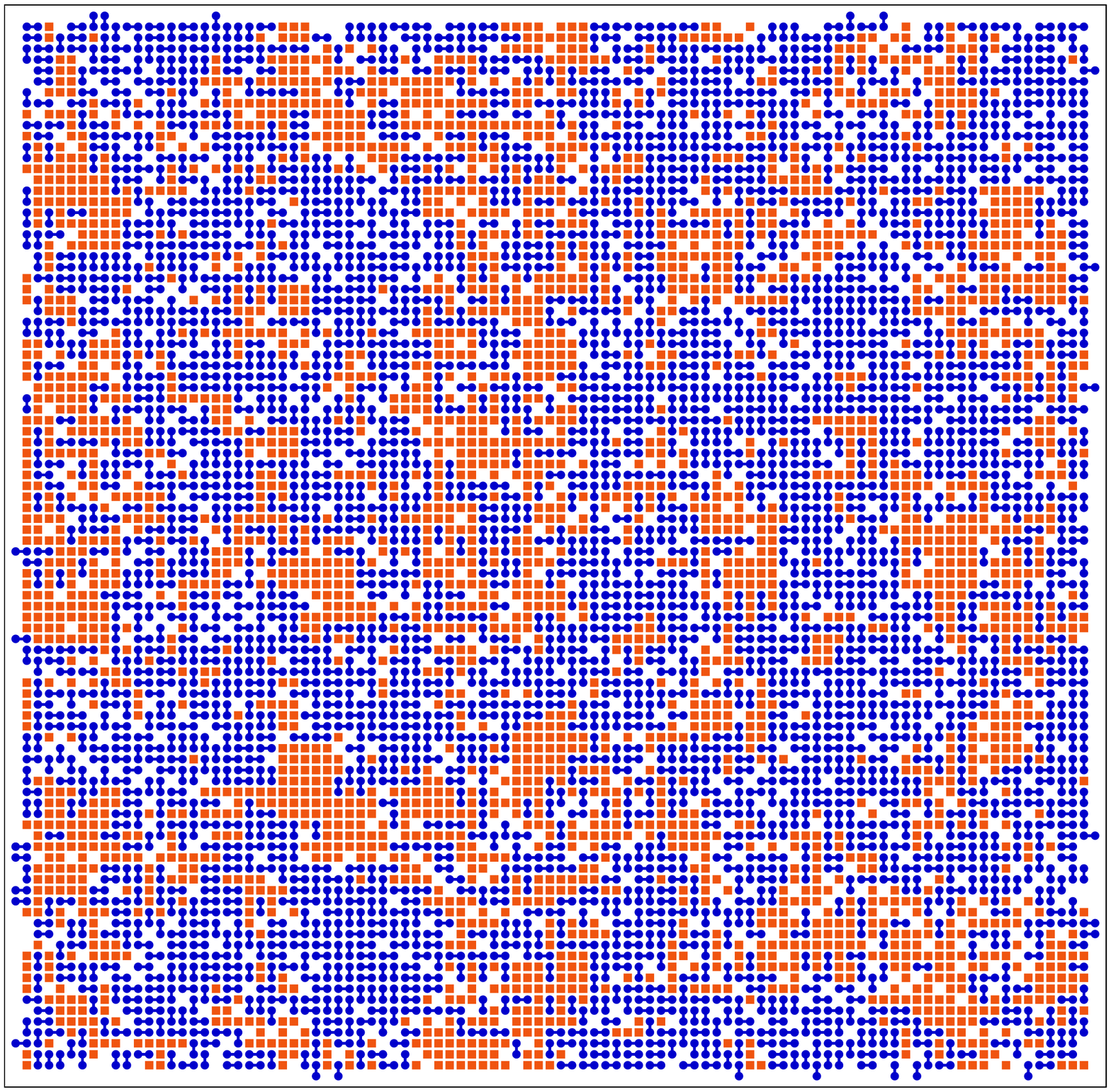}   \vspace*{0.3cm} \\
(a) & (b)
\end{tabular}
\caption{Typical jamming configurations of dimers on a $96 \times 96$ lattice with periodic boundary
   conditions using defect density $q = 0.3$ for (a) uncorrelated, and (b) correlated defects with 
   $\gamma = 0.4$. The sites with defects and the absorbed dimers have been painted in orange 
   squares and blue circles, respectively.}
\label{fig:jamconfig}
\end{figure}

\section{Results}

\subsection{Impact on the jamming coverage}

    In Fig. \ref{fig:jamconfig}, we have shown typical jamming configurations for a given defect 
    density $q$ for both correlated and uncorrelated defects. We first observe that the defects 
    form clusters via nearest neighbor connections, which become more and more compact for 
    stronger correlations (small $\gamma$). An idea of the compactness of the clusters may 
    quantitatively be realized from the fact that in the subcritical regime of defect density, 
    i.e., for $q < q_c$, where the global connectivity through the clusters of defects is absent, 
    the average size of the largest cluster of defects scales with $L$ as 
    $\langle s_{\rm max}^{\rm def}(L) \rangle \sim (\ln(L))^\alpha$ (not shown). It is found 
    that the exponent $\alpha > 1$ for $0 < \gamma < d$, which monotonically decreases with 
    increasing the value of $\gamma$. This finally approaches $\alpha = 1$, corresponding to 
    the value for uncorrelated defects. 
    
\begin{figure}[t]
\centering
\includegraphics[width=0.9\linewidth,trim={0.9cm 0.0cm 0.9cm 0.5cm},clip]{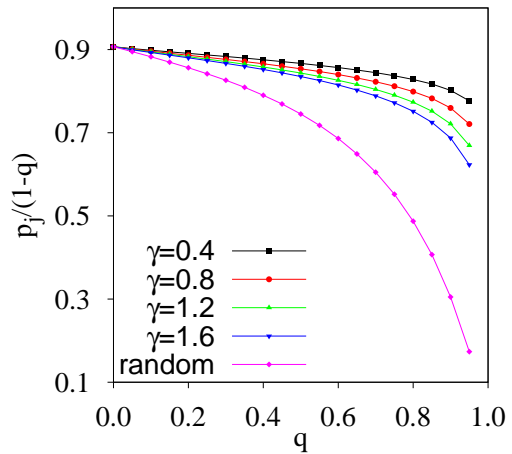}
\caption{The variation of filling fraction $p_j/(1-q)$ against the defect density 
   $q$ for $\gamma$ = 0.4, 0.8, 1.2, 1.6, and uncorrelated defects using  $L = 1024$.   
}
\label{fig:pj_vs_gamma}
\end{figure}

    Naturally, the particles experience the strongest screening effect in the case of homogeneously 
    distributed uncorrelated defects during the deposition, and it recedes as the strength of the
    correlation is increased. Thus, we expect to observe densely packed jamming states for stronger 
    correlations. To demonstrate this, the filling fraction $p_j/(1-q)$ at the jamming state is 
    plotted against $q$ in Fig.\ \ref{fig:pj_vs_gamma} for four different values of $\gamma$ and 
    for uncorrelated defects. Clearly, for any given value of $0 < q < 1$, the filling fraction 
    increases as the value of $\gamma$ decreases.
    
\subsection{Universality class of the jamming transition}

\begin{figure}[t]
\centering
\includegraphics[width=0.98\linewidth]{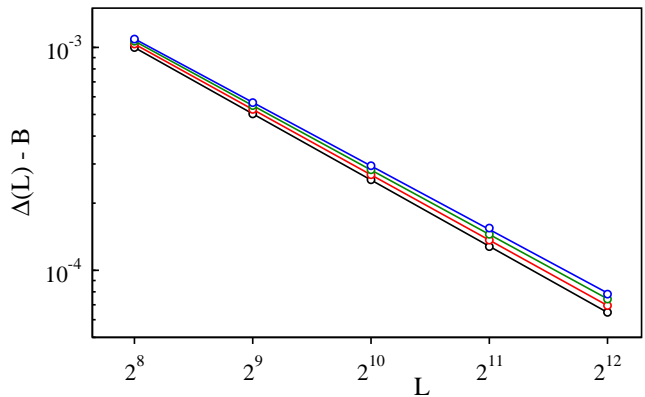}
\caption{Plot of $\Delta(L) - B$ against $L$ on a log-log scale for $\gamma=0.8$. The slope
   of the best fitted straight lines have been estimated as $1/\nu_j = 0.987(2), 0.974(4), 
   0.962(5),$ and $0.944(5)$ for $q$ = 0.1, 0.2, 0.3, and 0.4, respectively (arranged from 
   bottom to top). The corresponding values of $B$ are smaller than $B=9.77182 \times 10^{-6}$ for 
   $q=0.4$. 
}
\label{fig:pj_fluc}
\end{figure}

    To investigate whether the universality class of the jamming transition is affected 
    by the introduction of spatial correlations among the defects, we perform the scaling 
    analysis of the width $\Delta(L)$ of the transition zone. Precisely, we calculate the
    standard deviation of the jamming densities $\Delta(L) = \sqrt{\langle p_j^2 \rangle 
    - \langle p_j \rangle^2}$ for several values of $L$, which generally scales as $\Delta(L) 
    \sim L^{-1/\nu_j}$. Therefore, we plot $\Delta(L)$ versus $L$ on a log-log scale. We 
    simulate up to $L=4096$ and the averaging was done on (at least) $5 \times 10^6, 9.5 \times 
    10^5, 1.5 \times 10^5, 3 \times 10^4$, and $6 \times 10^3$ independent configurations 
    for $L$ = 256, 512, 1024, 2048, and 4096, respectively. It is observed that the curves 
    for small $\gamma$ have a certain amount of curvatures and they seem to approach a 
    constant value in the limit $L \to \infty$. We thus consider a modified functional 
    form
    \begin{equation}
        \Delta(L) = AL^{-1/\nu_j} + B
    \label{EQ:nu_j}    
    \end{equation}
    to fit our data. Indeed, a plot of $\Delta(L) - B$ against $L$ on a double logarithmic 
    scale exhibits a straight line, as shown in Fig.\ \ref{fig:pj_fluc}. It is found that 
    $B \approx 0$ for $\gamma \gtrsim 1.0$, but its value increases monotonically as
    $\gamma$ is decreased. To give an idea, the least-square fit of our data using Eq.\
    (\ref{EQ:nu_j}) yields the values of $(1/\nu_j, B) \approx (0.97, 4.63 \times 10^{-6}), 
    (0.95, 2.66 \times 10^{-5})$, and $(0.92, 9.31 \times 10^{-5})$ for $\gamma$ = 0.8, 
    0.6, and 0.4, respectively, using $q=0.2$. 
    As opposed to the case of uncorrelated defects, for which one obtains the universal 
    value of $\nu_j = 1$ in two dimensions at any arbitrary value of $0 < q < 1$, it is
    evident from Fig.\ \ref{fig:pj_fluc} that the exponent $\nu_j$ varies systematically 
    with $q$ for a fixed $\gamma$.
    
\begin{figure}[t]
\centering
\includegraphics[width=0.98\linewidth]{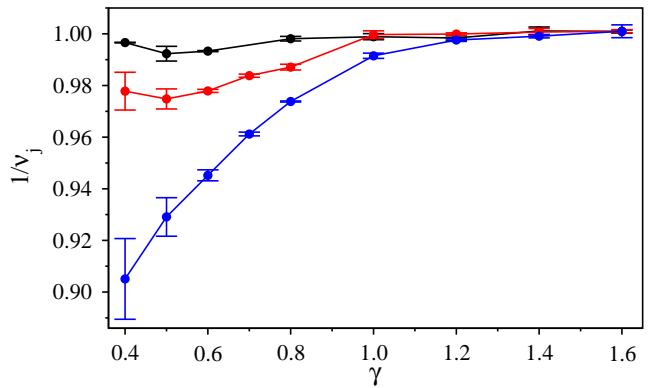}
\caption{Variation of the critical exponent $1/\nu_j$ associated with the jamming
   transition as a function of the correlation strength $\gamma$ for $q=0.05$ (black), 
   0.10 (red), and 0.20 (blue) (arranged from top to bottom).
}
\label{fig:nu_j}
\end{figure}

    Similar plots are made, but now we keep $q$ fixed and vary $\gamma$. To show the effect 
    of the spatial correlation more explicitly, we focus on the range of $q < q_c$, where 
    there exists no giant cluster of defects. It is found that the deviation of $\nu_j$ from 
    its universal value $\nu_j=1$ becomes more and more prominent as $\gamma \to 0$ (strong 
    correlations). In Fig.\ \ref{fig:nu_j}, we have displayed such a plot for three different 
    values of $q$. Although, the non-universal behavior is not so obvious from this figure 
    for $q=0.05$ as the exponent value is close to $\nu_j=1$, the curves have apparent 
    curvatures on the $\Delta(L)$ vs.\ $L$ plots for small $\gamma$. Precisely, the best fit 
    using Eq.\ (\ref{EQ:nu_j}) yields a value of $B > 0$ (e.g., $B \approx 2.1981 \times 
    10^{-5}$ for $\gamma=0.4$ and $q=0.05$). This suggests that a small but finite amount 
    of defects (i.e., $q>0$) is sufficient to change the universal behavior of the jamming 
    transition if and only if the defects have strong spatial correlations.
    
    What could be the reason behind the origin of this non-universal behavior? One may notice 
    from Fig.\ \ref{fig:jamconfig} that for strong spatial correlations, the void space becomes 
    fragmented into several isolated clusters and form islands surrounded by defects. This 
    happens even for a small value of $q$ (e.g., $q=0.1$), while they are less likely to be 
    formed in the case of uncorrelated defects at such small densities. The shapes and sizes of 
    these islands vary for different configurations. The size distribution appears to be broad 
    even for $q=0.1$, and the tail of the distribution is observed to shift to the origin with 
    increasing $\gamma$ (weak correlations). Besides that, the total number of islands also 
    varies for different configurations. Since the particle adsorption is occurring
    on these islands, one may think that the fluctuation of the jamming densities for a
    given system size $L$ is a collective contribution of the fluctuations arising from all 
    those islands. Thus, the variability of the island sizes could be the source 
    of breaking the universality class of the jamming transition. 

\begin{figure}[b]
\centering
\includegraphics[width=0.98\linewidth]{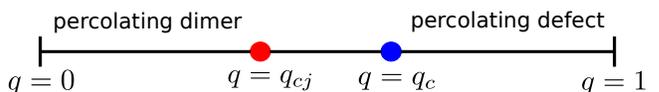}
\caption{Schematic phase diagram in one dimensional $q$ plane. Red and blue dots represent 
   the critical point of percolation through the dimers and defects respectively. In the 
   region between the two dots the system percolates neither through dimers nor through 
   defects.}
\label{fig:schematic}
\end{figure} 
    
    Arguably, such a scenario also arises at the percolation point of void spaces in the presence 
    of uncorrelated defects, where the size distribution of those islands follows a scale-free 
    distribution. However, using extensive numerical simulations by setting $1-q$ = 0.592746050 
    (percolation threshold of the square lattice), we have obtained the universal value of 
    $\nu_j=1$ (not shown). This suggests that the departure from the universal behavior for 
    long-range spatially correlated defects is probably not due to the above mentioned 
    fluctuations and could be related to some more complex details, such as spatial correlations 
    between the sizes of the islands or the non-trivial interactions of particles in the close 
    proximity to the complex inner and outer wall of the compact clusters of defects. 
    
    Furthermore, we have noticed that the distribution of $p_j$ deviates from a Gaussian 
    distribution for strongly correlated defects for large system sizes.

\begin{figure}
\centering
\includegraphics[width=0.98\linewidth]{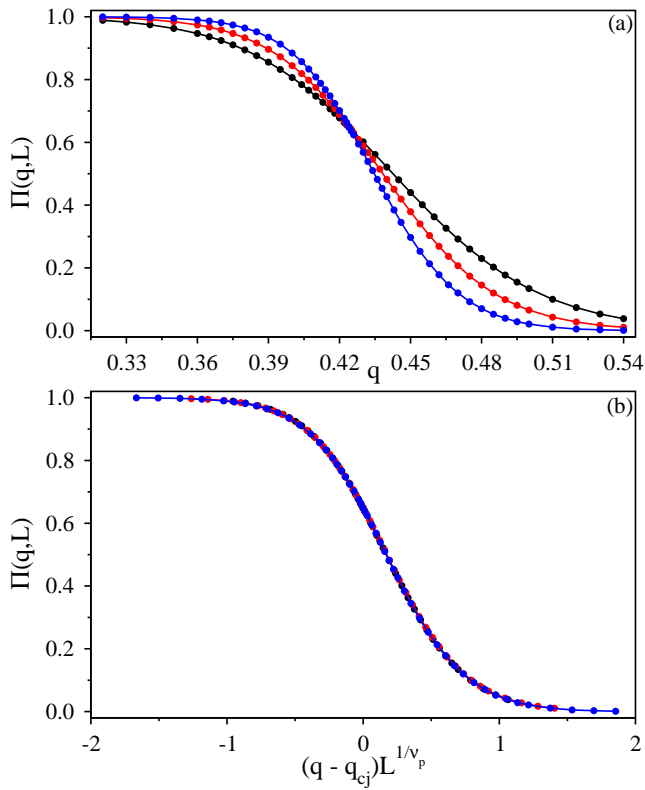}
\caption{For $\gamma=0.8$, (a) variation of the spanning probability $\Pi(q,L)$ of the 
   jamming configuration with defect density $q$ for system sizes $L$ = 256 (black), 
   512 (red) and 1024 (blue) (from left to right along $\Pi = 0.8$); (b) finite-size scaling 
   plot of the same data using $q_{cj} = 0.4241(2)$ and $1/\nu_p = 0.400(5)$.}
\label{fig:percprob0.8}
\end{figure}

\begin{figure}
\centering
\includegraphics[width=0.98\linewidth]{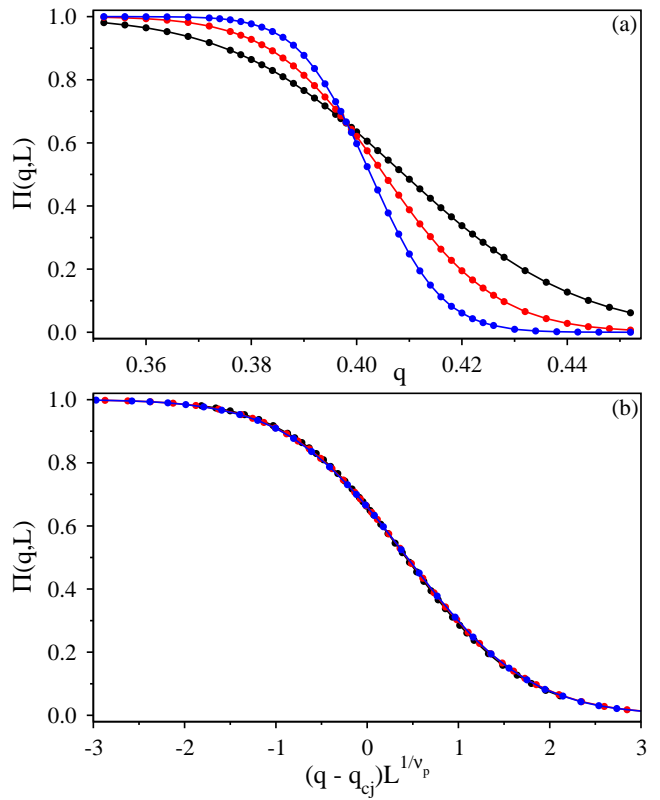}
\caption{For $\gamma=1.5$, (a) variation of the spanning probability $\Pi(q,L)$ of the 
   jamming configuration with defect density $q$ for system sizes $L$ = 256 (black), 
   512 (red) and 1024 (blue) (from left to right along $\Pi = 0.8$); (b) finite-size scaling 
   plot of the same data using $q_{cj} = 0.3982(2)$ and $1/\nu_p = 0.662(5)$.}
\label{fig:percprob1.5}
\end{figure}

\subsection{Percolation transition of the jamming states}

   We now identify the clusters of absorbed dimers in the jamming state, where a cluster 
   consists of a set of sites interconnected through their neighboring sites occupied by 
   the dimers. For $q=0$, the density of occupied sites is so high ($p_j \approx 0.9068$) 
   that there always exists global connectivity through a cluster spanning the entire 
   system. On the other hand, at $q=q_c$, when a giant cluster of defects first appears 
   in the system, the largest cluster of dimers becomes minuscule and it fails to establish 
   such global connectivity. Consequently, in between $q=0$ and $q_c$, one finds a 
   threshold value of $q=q_{cj}$ such that the system of dimers exhibits the global 
   connectivity and thus remains in the percolating phase only when $q<q_{cj}$. In the 
   range of $q_{cj} < q < q_c$, neither the largest cluster of dimers nor defects percolates. The schematic phase diagram in $q$ plane is shown in Fig.~\ref{fig:schematic}.
   
   The most important question here is, whether the critical behavior of such a percolation
   transition occurring at $q=q_{cj}$ belongs to the ordinary percolation universality 
   class when the defects have spatial long-range correlations. To investigate this, we 
   calculate the spanning probability $\Pi(q,L)$ that there exists a spanning cluster of 
   adsorbed dimers in the system by varying the value of $q$ for three different system 
   sizes $L$. Then, by performing the finite-size scaling analysis of $\Pi(q,L)$ and 
   estimating the scaling exponents we determine the universality class of the percolation 
   transition. Once a jamming configuration is reached in our simulation, we check the top 
   to bottom connectivity through the neighboring sites occupied by the dimers using the 
   Burning algorithm~\cite{Stauffer2018} imposing periodic (open) boundary conditions 
   along the horizontal (vertical) direction. It may be noted that the dimers adsorbed in 
   the isolated small islands of void space do not help in achieving the global connectivity. 
   Only the largest island of void space holds a special importance for this purpose, whose 
   size in its percolating phase scales as $\langle s_{max}^{\rm vac}(L) \rangle = 
   (a + b/\ln(L))L^2$ (not shown). In the context of percolation of adsorbed dimers, 
   this may signify an effective change in the dimensionality of the problem. In general, 
   $b>0$ for strong correlations, but its value decreases monotonically and approaches to 
   zero (the value in the case of uncorrelated defects) as $\gamma$ increases.
   
\begin{figure}[t]
\centering
\includegraphics[width=0.98\linewidth]{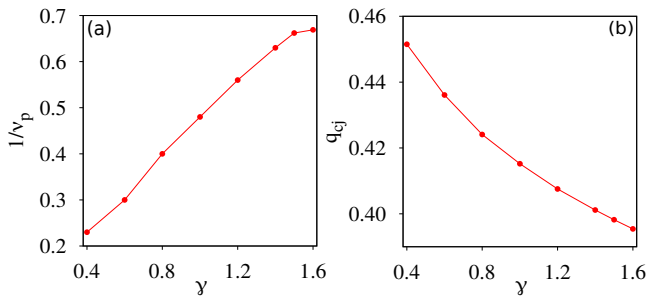}
\caption{Variation of (a) the critical exponent $\nu_p$ associated with the percolation
   transition of the jamming states and (b) the percolation threshold $q_{cj}$ as a 
   function of $\gamma$.  
}
\label{fig:gm_nu_qc}
\end{figure}

   In Fig.\ \ref{fig:percprob0.8}(a), the variation of the spanning probability $\Pi(q,L)$ 
   against $q$ has been shown for $\gamma=0.8$. By appropriately scaling the horizontal
   axis, when the same sets of data are re-plotted against $(q-q_{cj})L^{1/\nu_p}$ we 
   observe a nice data collapse [Fig.~\ref{fig:percprob0.8}(b)], implying the finite-size 
   scaling form
   \begin{equation}
       \Pi(q,L)\sim \mathcal{F}[(q-q_{cj})L^{1/\nu_p}],
   \end{equation}
   where $\nu_p$ is recognized as the correlation length exponent of the percolation transition.
   The analysis yields $q_{cj} = 0.4241(2)$ and $1/\nu_p = 0.400(5)$ for $\gamma=0.8$. We have 
   also shown similar plots for $\gamma=1.5$ in Figs.\ \ref{fig:percprob1.5} (a) and \ref{fig:percprob1.5}(b). In this 
   case we obtained $q_{cj} = 0.3982(2)$ and $1/\nu_p = 0.662(5)$. It is evident that these exponent 
   values are distinctly different from the value of $1/\nu_p = 0.75$ for uncorrelated defects, 
   for which such a transition belongs to the ordinary percolation universality class in two 
   dimensions. 
   
   Repeating these analyses for many different values of $\gamma$, we see that the critical 
   exponent $1/\nu_p$ increases with increasing $\gamma$ and approaches to $1/\nu_p=3/4$, as 
   shown in Fig.\ \ref{fig:gm_nu_qc}(a). This dependency is approximately described by a relation 
   $\nu_p = 2/\gamma$ in the range of $\gamma = 0.6-1.0$. The percolation threshold 
   $q_{cj}(\gamma)$ decreases with increasing $\gamma$ and approaches 0.3180(5), the 
   value for uncorrelated defects, as shown in Fig.\ \ref{fig:gm_nu_qc}(b). Note that our 
   results for uncorrelated defects are in good agreement with the previous numerical data 
   in Ref.\ \cite{Tarasevich2015}. The data used for all these plots are based on averages 
   over (at least) $10^6$, $5 \times 10^5$, and $7 \times 10^4$ samples for $L$ = 256, 512, 
   and 1024, respectively. Therefore, we believe that the above estimates are reasonably 
   accurate.

\subsection{Percolation transition before jamming}

   We have seen that for a given value of $\gamma$, the defect density $q=q_{cj}(\gamma)$ 
   separates between the percolating and nonpercolating jamming states. Specifically, all 
   the jamming configurations for $q<q_{cj}(\gamma)$ with density $p_j(\gamma,q)$ percolate 
   in the limit of asymptotically large system sizes. This suggests that for all values of 
   $q<q_{cj}(\gamma)$ there should be a critical value of $p=p_c(\gamma,q)$ such that the 
   system exhibits global connectivity for $p_c(\gamma,q) \leqslant p \leqslant p_j(\gamma,q)$.
   
\begin{figure}[t]
\centering
\includegraphics[width=0.98\linewidth]{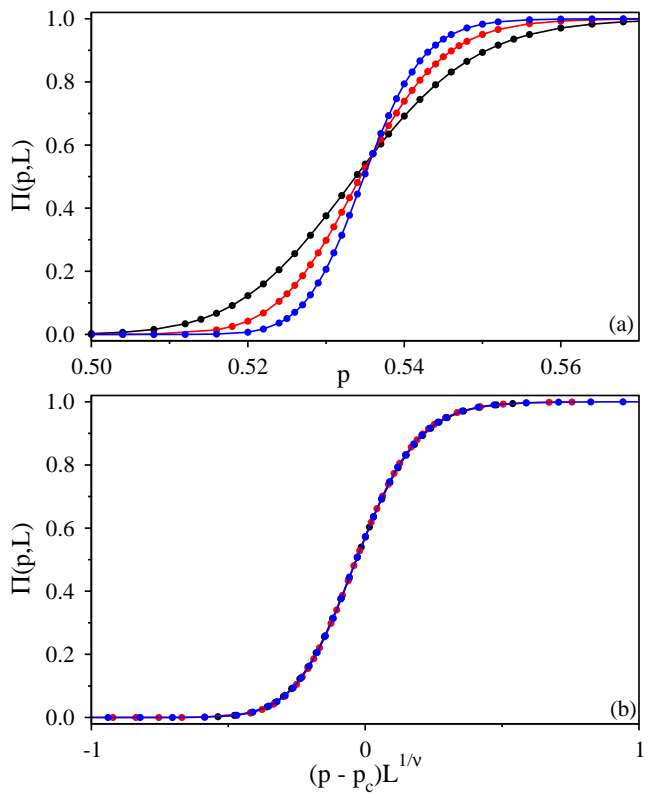}
\caption{For $\gamma = 0.8$, $q = 0.1$, (a) the spanning probability $\Pi(p,L)$ has been 
   plotted against the surface coverage $p$ for system sizes $L$ = 256 (black), 512 (red), 
   and 1024 (blue) (from left to right along $\Pi = 0.2$). (b) Finite-size scaling of the same data as in (a) using $p_c = 0.53595(3)$ 
   and $1/\nu = 0.488(5)$.}
\label{fig:percprob_q0.10}
\end{figure}

\begin{figure}[t]
\centering
\includegraphics[width=0.98\linewidth]{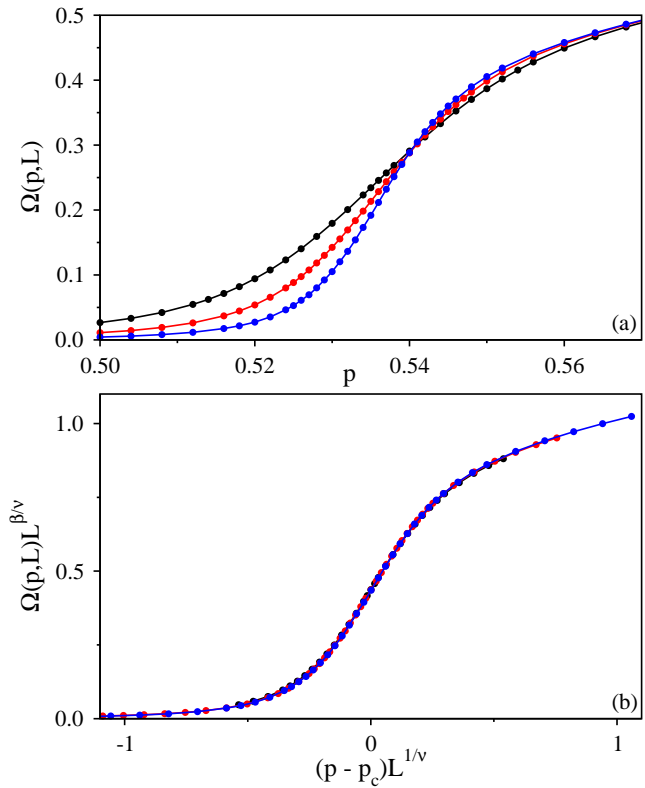}
\caption{For $\gamma = 0.8$, $q = 0.1$, (a) variation of the order parameter $\Omega(p,L)$ 
   against the surface coverage $p$ for system sizes $L$ = 256 (black), 512 (red), and 
   1024 (blue) (from left to right along $\Pi = 0.2$). (b) Finite-size scaling of the same data as 
   in (a) using $p_c = 0.53595(3)$, $1/\nu = 0.488(5)$, and $\beta/\nu = 0.104(1)$.}
\label{fig:orderpara_q0.10}
\end{figure}

   In Fig.\ \ref{fig:percprob_q0.10}(a), we have plotted the spanning probability $\Pi(p,L)$ 
   against the surface coverage $p$ for three different system sizes using $\gamma = 0.8$ 
   and $q = 0.1$. These curves cross each other approximately at a single point 
   $[p_c(\gamma,q),\Pi(p_c)]$. From visual inspection, we estimate that 
   $p_c(0.8,0.1)=0.53595(3)$ and $\Pi(p_c) \approx 0.57$, which is quite lower than the 
   value 0.\numprint{636454001} of the crossing probability on a cylindrical geometry 
   obtained using Cardy's formula~\cite{Cardy2006,Ziff2011} for defect-free systems.
   It is to be noted that the crossing probability $\Pi(p_c) \approx 0.64$ has been 
   obtained even for the system with uncorrelated defects.
   Now a finite-size scaling of the same data is performed. A plot of $\Pi(p,L)$ against the scaled 
   variable $(p - p_c(\gamma,q))L^{1/\nu}$ exhibits the data collapse for all three system sizes 
   [Fig.\ \ref{fig:percprob_q0.10}(b)], implying the scaling form
   \begin{equation}
   \Pi(p,L) \sim \mathcal{F}[(p - p_c(\gamma,q))L^{1/\nu}].  
   \end{equation}
   In percolation problems, the average size of the largest cluster per site is considered as 
   the order parameter $\Omega(p,L) = \langle s_{max}(p,L) \rangle / L^2$, where $s_{max}$ 
   represents the size of the largest cluster of absorbed dimers. In Fig.\ 
   \ref{fig:orderpara_q0.10}(a), we have shown the variation of $\Omega(p,L)$ against $p$ 
   for the same three system sizes. Again, by appropriately scaling the abscissa and 
   ordinate, and re-plotting the data we observe data collapse of $\Omega(p,L)$, as shown 
   in Fig.\ \ref{fig:orderpara_q0.10}(b), indicating the scaling form
   \begin{equation}
   \Omega(p,L)L^{\beta/\nu} \sim \mathcal{G}[(p - p_c(\gamma,q))L^{1/\nu}].  
   \end{equation}
   The finite-size scaling analysis yields $p_c(0.8,0.1) = 0.53595(3)$, $1/\nu = 0.488(5)$, 
   and $\beta/\nu = 0.104(1)$. These values are compared with the known critical exponents 
   for ordinary percolation in two dimensions, which are $1/\nu = 3/4$ and $\beta = 5/36$.
   
\begin{table}
   \caption{Our numerical estimates of the percolation threshold $p_c(\gamma,q)$ 
   for different values of the defect density $q$ and the correlation strength
   $\gamma$. The numbers in the parentheses represent the error bars in the last 
   digit.}
\begin{tabular*}{\linewidth}{c@{\extracolsep{\fill}}cccr}
\hline \hline \vspace{-0.27cm} \\
  $\gamma$  & $q$    & $p_c(\gamma,q)$       & $1/\nu$  & $\beta/\nu$   \\ \vspace{-0.27cm} \\ \hline \hline
    0.6     & 0.01   & 0.55965(3)  & 0.71(1) & 0.105(1) \\ \hline
            & 0.01   & 0.56022(3)  & 0.718(5) & 0.104(1) \\
    0.8     & 0.05   & 0.54986(3)  & 0.587(5) & 0.104(1) \\
            & 0.10   & 0.53595(3)  & 0.488(6) & 0.104(1) \\ \hline
            & 0.01   & 0.56083(2)  & 0.735(5) & 0.104(1) \\
    1.2     & 0.05   & 0.55362(2)  & 0.671(6) & 0.105(1) \\
            & 0.10   & 0.54329(3)  & 0.621(5) & 0.104(1)  \\ 
    \hline \hline
\end{tabular*}
\label{tab:exponents}
\end{table}
   
   The above set of calculations has been repeated for different $(\gamma,q)$ pairs.
   Interestingly, we found that the critical exponent $1/\nu$ depends systematically 
   with $q$ and $\gamma$, whereas $\beta/\nu$ always appears to be same as the value 
   of the ordinary percolation in two dimensions, i.e., $\beta/\nu=5/48$. The 
   estimated values for different $(\gamma,q)$ pairs are listed in Table 
   \ref{tab:exponents}. For strong correlations ($\gamma \leqslant 0.6$) and high $q$, 
   it is observed that the crossing points of the curves for $\Pi(p,L)$ vary over a 
   much wider range. In these cases, the two-parameter scaling plot does not exhibit 
   an excellent data collapse as seen for $\gamma \geqslant 0.8$. Probably, 
   logarithmic corrections are responsible for this. Further investigations using 
   higher system sizes are thus needed for a precise understanding of this problem.


\section{Conclusions}

   We have investigated the percolation and jamming properties of the random sequential 
   adsorption of dimers on square
   lattice in the presence of defects with spatial long-range correlations. Accordingly,
   a fraction $q$ of the lattice sites are declared as defects, where the deposition
   of dimers is completely forbidden. The dimer adsorption takes place randomly at 
   the available vacant space. The correlation strength among the defective sites is 
   varied and its impact on the jamming and percolation transitions has been studied 
   using extensive numerical simulations.
   
   It has been observed that the jamming coverage for any arbitrary value of $0<q<1$
   is increased with increasing correlation strength. More importantly, for 
   strong correlations, the jamming transition is found to be non-universal even 
   when $q$ is much smaller than its threshold value $q_c$ such that the connected 
   clusters of defects are all minuscule. A continuously tunable value of $\nu_j$ 
   characterizes the jamming transition, which approaches its universal value 
   $\nu_j = 1$ in two dimensions with decreasing correlation strength.

   The percolation transition of the absorbed dimers takes place at a critical density of
   occupied sites $p_c$ only when the defect density is smaller than a critical value 
   $q = q_{cj}$. The percolation threshold $p_c$ has been found to be dependent on the 
   defect density as well as the strength of the spatial correlations. For a given defect
   density, $p_c$ decreases as the correlation strength is increased. Moreover, the 
   finite-size scaling analysis reveals that the transition does not belong to the 
   ordinary percolation universality class. The correlation length exponent $\nu$ 
   changes systematically with the strength of the spatial correlation and approaches 
   to its universal value $4/3$ in two dimensions when the defects become weakly 
   correlated. Remarkably, the ratio of the exponents $\beta/\nu$ associated with the 
   order parameter scaling appears to remain the same as the ordinary percolation.
   
   Finally, by tuning the defect density $q$ a percolation transition is observed at 
   $q=q_{cj}$, which separates the percolating jamming states from the non-percolating 
   ones. Again, the percolation transition is characterized by a non-universal value 
   of the correlation length exponent, which is found to be dependent on the strength 
   of the spatial correlation among the defects.
   
   In the future, apart from the obvious generalization of this model by considering 
   different shapes and sizes of the particles on different lattice geometries or 
   in higher dimensions, one may find it interesting to study precursor-mediated 
   adsorption in such a correlated disordered environment.
   
   We are hopeful that the results presented here will provide a framework for 
   understanding various observations in different experimental conditions 
   more coherently, since by tuning the parameters ($q$ and $\gamma$) of the model 
   a system resembling the real one may be devised.
   
\section*{Acknowledgements}

   We sincerely thank Nuno A. M. Ara\'{u}jo for a critical review of the manuscript 
   and A. J. Ramirez-Pastor for suggesting to us interesting papers on the subject.
   
   S. K. acknowledges support from the Japan Society for the Promotion of Science (JSPS) 
   Grants-in-Aid for Scientific Research (KAKENHI) Grants Nos. JP16H06478 and 19H01811. 
   Additional support from the MEXT under “Exploratory Challenge on Post-K computer” 
   (Frontiers of Basic Science: Challenging the Limits) and the “Earthquake and Volcano 
   Hazards Observation and Research Program” is also gratefully acknowledged.
   
   D.M. acknowledges the support from a EPSRC Programme Grant (Grant EP/R018820/1) which funds the Crystallization in the Real World consortium. In addition, D.M. gratefully acknowledges the use of the computational facilities provided by the University of Warwick Centre for Scientific Computing.



\begin{thebibliography}{43}%
\makeatletter
\providecommand \@ifxundefined [1]{%
 \@ifx{#1\undefined}
}%
\providecommand \@ifnum [1]{%
 \ifnum #1\expandafter \@firstoftwo
 \else \expandafter \@secondoftwo
 \fi
}%
\providecommand \@ifx [1]{%
 \ifx #1\expandafter \@firstoftwo
 \else \expandafter \@secondoftwo
 \fi
}%
\providecommand \natexlab [1]{#1}%
\providecommand \enquote  [1]{``#1''}%
\providecommand \bibnamefont  [1]{#1}%
\providecommand \bibfnamefont [1]{#1}%
\providecommand \citenamefont [1]{#1}%
\providecommand \href@noop [0]{\@secondoftwo}%
\providecommand \href [0]{\begingroup \@sanitize@url \@href}%
\providecommand \@href[1]{\@@startlink{#1}\@@href}%
\providecommand \@@href[1]{\endgroup#1\@@endlink}%
\providecommand \@sanitize@url [0]{\catcode `\\12\catcode `\$12\catcode
  `\&12\catcode `\#12\catcode `\^12\catcode `\_12\catcode `\%12\relax}%
\providecommand \@@startlink[1]{}%
\providecommand \@@endlink[0]{}%
\providecommand \url  [0]{\begingroup\@sanitize@url \@url }%
\providecommand \@url [1]{\endgroup\@href {#1}{\urlprefix }}%
\providecommand \urlprefix  [0]{URL }%
\providecommand \Eprint [0]{\href }%
\providecommand \doibase [0]{http://dx.doi.org/}%
\providecommand \selectlanguage [0]{\@gobble}%
\providecommand \bibinfo  [0]{\@secondoftwo}%
\providecommand \bibfield  [0]{\@secondoftwo}%
\providecommand \translation [1]{[#1]}%
\providecommand \BibitemOpen [0]{}%
\providecommand \bibitemStop [0]{}%
\providecommand \bibitemNoStop [0]{.\EOS\space}%
\providecommand \EOS [0]{\spacefactor3000\relax}%
\providecommand \BibitemShut  [1]{\csname bibitem#1\endcsname}%
\let\auto@bib@innerbib\@empty
\bibitem [{\citenamefont {Evans}(1993)}]{Evans1993}%
  \BibitemOpen
  \bibfield  {author} {\bibinfo {author} {\bibfnamefont {J.~W.}\ \bibnamefont
  {Evans}},\ }\href@noop {} {\bibfield  {journal} {\bibinfo  {journal} {Rev.
  Mod. Phys.}\ }\textbf {\bibinfo {volume} {65}},\ \bibinfo {pages} {1281}
  (\bibinfo {year} {1993})}\BibitemShut {NoStop}%
\bibitem [{\citenamefont {Feder}(1980)}]{Feder1980}%
  \BibitemOpen
  \bibfield  {author} {\bibinfo {author} {\bibfnamefont {J.}~\bibnamefont
  {Feder}},\ }\href@noop {} {\bibfield  {journal} {\bibinfo  {journal} {Journal
  of Theoretical Biology}\ }\textbf {\bibinfo {volume} {87}},\ \bibinfo {pages}
  {237 } (\bibinfo {year} {1980})}\BibitemShut {NoStop}%
\bibitem [{\citenamefont {Torquato}\ and\ \citenamefont
  {Stillinger}(2010)}]{Torquato2010}%
  \BibitemOpen
  \bibfield  {author} {\bibinfo {author} {\bibfnamefont {S.}~\bibnamefont
  {Torquato}}\ and\ \bibinfo {author} {\bibfnamefont {F.~H.}\ \bibnamefont
  {Stillinger}},\ }\href@noop {} {\bibfield  {journal} {\bibinfo  {journal}
  {Rev. Mod. Phys.}\ }\textbf {\bibinfo {volume} {82}},\ \bibinfo {pages}
  {2633} (\bibinfo {year} {2010})}\BibitemShut {NoStop}%
\bibitem [{\citenamefont {Kumacheva}\ \emph {et~al.}(2002)\citenamefont
  {Kumacheva}, \citenamefont {Golding}, \citenamefont {Allard},\ and\
  \citenamefont {Sargent}}]{Kumacheva2002}%
  \BibitemOpen
  \bibfield  {author} {\bibinfo {author} {\bibfnamefont {E.}~\bibnamefont
  {Kumacheva}}, \bibinfo {author} {\bibfnamefont {R.}~\bibnamefont {Golding}},
  \bibinfo {author} {\bibfnamefont {M.}~\bibnamefont {Allard}}, \ and\ \bibinfo
  {author} {\bibfnamefont {E.}~\bibnamefont {Sargent}},\ }\href@noop {}
  {\bibfield  {journal} {\bibinfo  {journal} {Advanced Materials}\ }\textbf
  {\bibinfo {volume} {14}},\ \bibinfo {pages} {221} (\bibinfo {year}
  {2002})}\BibitemShut {NoStop}%
\bibitem [{\citenamefont {Hlady}\ and\ \citenamefont
  {Buijs}(1996)}]{Hlady1996}%
  \BibitemOpen
  \bibfield  {author} {\bibinfo {author} {\bibfnamefont {V.}~\bibnamefont
  {Hlady}, \bibfnamefont {V}}\ and\ \bibinfo {author} {\bibfnamefont
  {J.}~\bibnamefont {Buijs}},\ }\href@noop {} {\bibfield  {journal} {\bibinfo
  {journal} {Current opinion in biotechnology}\ }\textbf {\bibinfo {volume}
  {7}},\ \bibinfo {pages} {72} (\bibinfo {year} {1996})}\BibitemShut {NoStop}%
\bibitem [{\citenamefont {Roman}\ and\ \citenamefont
  {Majlis}(1983)}]{Roman1983}%
  \BibitemOpen
  \bibfield  {author} {\bibinfo {author} {\bibfnamefont {E.}~\bibnamefont
  {Roman}}\ and\ \bibinfo {author} {\bibfnamefont {N.}~\bibnamefont {Majlis}},\
  }\href@noop {} {\bibfield  {journal} {\bibinfo  {journal} {Solid State
  Communications}\ }\textbf {\bibinfo {volume} {47}},\ \bibinfo {pages} {259 }
  (\bibinfo {year} {1983})}\BibitemShut {NoStop}%
\bibitem [{\citenamefont {Yu}\ \emph {et~al.}(2016)\citenamefont {Yu},
  \citenamefont {Yang}, \citenamefont {Chen}, \citenamefont {Tao},\ and\
  \citenamefont {Liu}}]{Yu2016}%
  \BibitemOpen
  \bibfield  {author} {\bibinfo {author} {\bibfnamefont {D.}~\bibnamefont
  {Yu}}, \bibinfo {author} {\bibfnamefont {Y.-Q.}\ \bibnamefont {Yang}},
  \bibinfo {author} {\bibfnamefont {Z.}~\bibnamefont {Chen}}, \bibinfo {author}
  {\bibfnamefont {Y.}~\bibnamefont {Tao}}, \ and\ \bibinfo {author}
  {\bibfnamefont {Y.-F.}\ \bibnamefont {Liu}},\ }\href@noop {} {\bibfield
  {journal} {\bibinfo  {journal} {Optics Communications}\ }\textbf {\bibinfo
  {volume} {362}},\ \bibinfo {pages} {43 } (\bibinfo {year} {2016})},\ \bibinfo
  {note} {polymer Photonics and Its Applications}\BibitemShut {NoStop}%
\bibitem [{\citenamefont {Privman}(1994)}]{Privman1994}%
  \BibitemOpen
  \bibfield  {author} {\bibinfo {author} {\bibfnamefont {V.}~\bibnamefont
  {Privman}},\ }\href@noop {} {\bibfield  {journal} {\bibinfo  {journal}
  {Trends in Stat. Phys.}\ }\textbf {\bibinfo {volume} {1}},\ \bibinfo {pages}
  {89} (\bibinfo {year} {1994})}\BibitemShut {NoStop}%
\bibitem [{\citenamefont {Ramsden}(1992)}]{Ramsden1992}%
  \BibitemOpen
  \bibfield  {author} {\bibinfo {author} {\bibfnamefont {J.~J.}\ \bibnamefont
  {Ramsden}},\ }\href@noop {} {\bibfield  {journal} {\bibinfo  {journal} {The
  Journal of Physical Chemistry}\ }\textbf {\bibinfo {volume} {96}},\ \bibinfo
  {pages} {3388} (\bibinfo {year} {1992})}\BibitemShut {NoStop}%
\bibitem [{\citenamefont {Joshi}\ \emph {et~al.}(2016)\citenamefont {Joshi},
  \citenamefont {Bargteil}, \citenamefont {Caciagli}, \citenamefont
  {Burelbach}, \citenamefont {Xing}, \citenamefont {Nunes}, \citenamefont
  {Pinto}, \citenamefont {Ara{\'{u}}jo}, \citenamefont {Brujic},\ and\
  \citenamefont {Eiser}}]{Joshi2016}%
  \BibitemOpen
  \bibfield  {author} {\bibinfo {author} {\bibfnamefont {D.}~\bibnamefont
  {Joshi}}, \bibinfo {author} {\bibfnamefont {D.}~\bibnamefont {Bargteil}},
  \bibinfo {author} {\bibfnamefont {A.}~\bibnamefont {Caciagli}}, \bibinfo
  {author} {\bibfnamefont {J.}~\bibnamefont {Burelbach}}, \bibinfo {author}
  {\bibfnamefont {Z.}~\bibnamefont {Xing}}, \bibinfo {author} {\bibfnamefont
  {A.~S.}\ \bibnamefont {Nunes}}, \bibinfo {author} {\bibfnamefont {D.~E.}\
  \bibnamefont {Pinto}}, \bibinfo {author} {\bibfnamefont {N.~A.}\ \bibnamefont
  {Ara{\'{u}}jo}}, \bibinfo {author} {\bibfnamefont {J.}~\bibnamefont
  {Brujic}}, \ and\ \bibinfo {author} {\bibfnamefont {E.}~\bibnamefont
  {Eiser}},\ }\href@noop {} {\bibfield  {journal} {\bibinfo  {journal} {Science
  Advances}\ } (\bibinfo {year} {2016})},\ \Eprint
  {http://arxiv.org/abs/1603.05931} {1603.05931} \BibitemShut {NoStop}%
\bibitem [{\citenamefont {Napolitano}(2020)}]{Napolitano2020}%
  \BibitemOpen
  \bibfield  {author} {\bibinfo {author} {\bibfnamefont {S.}~\bibnamefont
  {Napolitano}},\ }\href@noop {} {\bibfield  {journal} {\bibinfo  {journal}
  {Soft Matter}\ }\textbf {\bibinfo {volume} {16}},\ \bibinfo {pages} {5348}
  (\bibinfo {year} {2020})}\BibitemShut {NoStop}%
\bibitem [{\citenamefont {Privman}(2000)}]{Privman2000}%
  \BibitemOpen
  \bibfield  {author} {\bibinfo {author} {\bibfnamefont {V.}~\bibnamefont
  {Privman}},\ }\href@noop {} {\bibfield  {journal} {\bibinfo  {journal} {The
  Journal of Adhesion}\ }\textbf {\bibinfo {volume} {74}},\ \bibinfo {pages}
  {421} (\bibinfo {year} {2000})}\BibitemShut {NoStop}%
\bibitem [{\citenamefont {Cadilhe}\ \emph {et~al.}(2007)\citenamefont
  {Cadilhe}, \citenamefont {Ara{\'{u}}jo},\ and\ \citenamefont
  {Privman}}]{Cadilhe2007}%
  \BibitemOpen
  \bibfield  {author} {\bibinfo {author} {\bibfnamefont {A.}~\bibnamefont
  {Cadilhe}}, \bibinfo {author} {\bibfnamefont {N.~A.~M.}\ \bibnamefont
  {Ara{\'{u}}jo}}, \ and\ \bibinfo {author} {\bibfnamefont {V.}~\bibnamefont
  {Privman}},\ }\href@noop {} {\bibfield  {journal} {\bibinfo  {journal}
  {Journal of Physics: Condensed Matter}\ }\textbf {\bibinfo {volume} {19}},\
  \bibinfo {pages} {065124} (\bibinfo {year} {2007})}\BibitemShut {NoStop}%
\bibitem [{\citenamefont {Flory}(1939)}]{Flory1939}%
  \BibitemOpen
  \bibfield  {author} {\bibinfo {author} {\bibfnamefont {P.~J.}\ \bibnamefont
  {Flory}},\ }\href@noop {} {\bibfield  {journal} {\bibinfo  {journal} {Journal
  of the American Chemical Society}\ }\textbf {\bibinfo {volume} {61}},\
  \bibinfo {pages} {1518} (\bibinfo {year} {1939})}\BibitemShut {NoStop}%
\bibitem [{\citenamefont {R\'{e}nyi}(1958)}]{Renyi1958}%
  \BibitemOpen
  \bibfield  {author} {\bibinfo {author} {\bibfnamefont {A.}~\bibnamefont
  {R\'{e}nyi}},\ }\href@noop {} {\bibfield  {journal} {\bibinfo  {journal}
  {Publ. Math. Inst. Hung. Acad. Sci}\ }\textbf {\bibinfo {volume} {3}},\
  \bibinfo {pages} {109} (\bibinfo {year} {1958})}\BibitemShut {NoStop}%
\bibitem [{\citenamefont {Kondrat}\ and\ \citenamefont
  {P\ifmmode~\mbox{\c{e}}\else \c{e}\fi{}kalski}(2001)}]{Kondrat2001}%
  \BibitemOpen
  \bibfield  {author} {\bibinfo {author} {\bibfnamefont {G.}~\bibnamefont
  {Kondrat}}\ and\ \bibinfo {author} {\bibfnamefont {A.}~\bibnamefont
  {P\ifmmode~\mbox{\c{e}}\else \c{e}\fi{}kalski}},\ }\href@noop {} {\bibfield
  {journal} {\bibinfo  {journal} {Phys. Rev. E}\ }\textbf {\bibinfo {volume}
  {63}},\ \bibinfo {pages} {051108} (\bibinfo {year} {2001})}\BibitemShut
  {NoStop}%
\bibitem [{\citenamefont {Cornette}\ \emph {et~al.}(2003)\citenamefont
  {Cornette}, \citenamefont {Ramirez-Pastor},\ and\ \citenamefont
  {Nieto}}]{Cornette2003}%
  \BibitemOpen
  \bibfield  {author} {\bibinfo {author} {\bibfnamefont {V.}~\bibnamefont
  {Cornette}}, \bibinfo {author} {\bibfnamefont {A.~J.}\ \bibnamefont
  {Ramirez-Pastor}}, \ and\ \bibinfo {author} {\bibfnamefont {F.}~\bibnamefont
  {Nieto}},\ }\href@noop {} {\bibfield  {journal} {\bibinfo  {journal} {Eur.
  Phys. J. B}\ }\textbf {\bibinfo {volume} {36}},\ \bibinfo {pages} {391}
  (\bibinfo {year} {2003})}\BibitemShut {NoStop}%
\bibitem [{\citenamefont {Tarasevich}\ \emph {et~al.}(2012)\citenamefont
  {Tarasevich}, \citenamefont {Lebovka},\ and\ \citenamefont
  {Laptev}}]{Tarasevich2012}%
  \BibitemOpen
  \bibfield  {author} {\bibinfo {author} {\bibfnamefont {Y.~Y.}\ \bibnamefont
  {Tarasevich}}, \bibinfo {author} {\bibfnamefont {N.~I.}\ \bibnamefont
  {Lebovka}}, \ and\ \bibinfo {author} {\bibfnamefont {V.~V.}\ \bibnamefont
  {Laptev}},\ }\href@noop {} {\bibfield  {journal} {\bibinfo  {journal} {Phys.
  Rev. E}\ }\textbf {\bibinfo {volume} {86}},\ \bibinfo {pages} {061116}
  (\bibinfo {year} {2012})}\BibitemShut {NoStop}%
\bibitem [{\citenamefont {Gimenez}\ and\ \citenamefont
  {Ramirez-Pastor}(2015)}]{Gimenez2015}%
  \BibitemOpen
  \bibfield  {author} {\bibinfo {author} {\bibfnamefont {M.}~\bibnamefont
  {Gimenez}}\ and\ \bibinfo {author} {\bibfnamefont {A.}~\bibnamefont
  {Ramirez-Pastor}},\ }\href@noop {} {\bibfield  {journal} {\bibinfo  {journal}
  {Physica A: Statistical Mechanics and its Applications}\ }\textbf {\bibinfo
  {volume} {421}},\ \bibinfo {pages} {261 } (\bibinfo {year}
  {2015})}\BibitemShut {NoStop}%
\bibitem [{\citenamefont {Stauffer}\ and\ \citenamefont
  {Aharony}(2018)}]{Stauffer2018}%
  \BibitemOpen
  \bibfield  {author} {\bibinfo {author} {\bibfnamefont {D.}~\bibnamefont
  {Stauffer}}\ and\ \bibinfo {author} {\bibfnamefont {A.}~\bibnamefont
  {Aharony}},\ }\href@noop {} {\emph {\bibinfo {title} {Introduction To
  Percolation Theory}}}\ (\bibinfo  {publisher} {Taylor {\&} Francis},\
  \bibinfo {year} {2018})\BibitemShut {NoStop}%
\bibitem [{\citenamefont {Becklehimer}\ and\ \citenamefont
  {Pandey}(1992)}]{Becklehimer1992}%
  \BibitemOpen
  \bibfield  {author} {\bibinfo {author} {\bibfnamefont {J.}~\bibnamefont
  {Becklehimer}}\ and\ \bibinfo {author} {\bibfnamefont {R.~B.}\ \bibnamefont
  {Pandey}},\ }\href@noop {} {\bibfield  {journal} {\bibinfo  {journal}
  {Physica A: Statistical Mechanics and its Applications}\ }\textbf {\bibinfo
  {volume} {187}},\ \bibinfo {pages} {71 } (\bibinfo {year}
  {1992})}\BibitemShut {NoStop}%
\bibitem [{\citenamefont {Lebovka}\ \emph {et~al.}(2011)\citenamefont
  {Lebovka}, \citenamefont {Karmazina}, \citenamefont {Tarasevich},\ and\
  \citenamefont {Laptev}}]{Lebovka2011}%
  \BibitemOpen
  \bibfield  {author} {\bibinfo {author} {\bibfnamefont {N.~I.}\ \bibnamefont
  {Lebovka}}, \bibinfo {author} {\bibfnamefont {N.~N.}\ \bibnamefont
  {Karmazina}}, \bibinfo {author} {\bibfnamefont {Y.~Y.}\ \bibnamefont
  {Tarasevich}}, \ and\ \bibinfo {author} {\bibfnamefont {V.~V.}\ \bibnamefont
  {Laptev}},\ }\href@noop {} {\bibfield  {journal} {\bibinfo  {journal} {Phys.
  Rev. E}\ }\textbf {\bibinfo {volume} {84}},\ \bibinfo {pages} {061603}
  (\bibinfo {year} {2011})}\BibitemShut {NoStop}%
\bibitem [{\citenamefont {Cie\'{s}\l{}a}\ \emph {et~al.}(2016)\citenamefont
  {Cie\'{s}\l{}a}, \citenamefont {Paj\c{a}k},\ and\ \citenamefont
  {Ziff}}]{ziff2016}%
  \BibitemOpen
  \bibfield  {author} {\bibinfo {author} {\bibfnamefont {M.}~\bibnamefont
  {Cie\'{s}\l{}a}}, \bibinfo {author} {\bibfnamefont {G.}~\bibnamefont
  {Paj\c{a}k}}, \ and\ \bibinfo {author} {\bibfnamefont {R.~M.}\ \bibnamefont
  {Ziff}},\ }\href@noop {} {\bibfield  {journal} {\bibinfo  {journal} {The
  Journal of Chemical Physics}\ }\textbf {\bibinfo {volume} {145}},\ \bibinfo
  {pages} {044708} (\bibinfo {year} {2016})}\BibitemShut {NoStop}%
\bibitem [{\citenamefont {Guo}\ \emph {et~al.}(1994)\citenamefont {Guo},
  \citenamefont {Bradley}, \citenamefont {Hopkinson},\ and\ \citenamefont
  {King}}]{Guo1994}%
  \BibitemOpen
  \bibfield  {author} {\bibinfo {author} {\bibfnamefont {X.-C.}\ \bibnamefont
  {Guo}}, \bibinfo {author} {\bibfnamefont {J.}~\bibnamefont {Bradley}},
  \bibinfo {author} {\bibfnamefont {A.}~\bibnamefont {Hopkinson}}, \ and\
  \bibinfo {author} {\bibfnamefont {D.}~\bibnamefont {King}},\ }\href@noop {}
  {\bibfield  {journal} {\bibinfo  {journal} {Surface Science}\ }\textbf
  {\bibinfo {volume} {310}},\ \bibinfo {pages} {163 } (\bibinfo {year}
  {1994})}\BibitemShut {NoStop}%
\bibitem [{\citenamefont {Kundu}\ \emph {et~al.}(2018)\citenamefont {Kundu},
  \citenamefont {Ara\'ujo},\ and\ \citenamefont {Manna}}]{Kundu2018}%
  \BibitemOpen
  \bibfield  {author} {\bibinfo {author} {\bibfnamefont {S.}~\bibnamefont
  {Kundu}}, \bibinfo {author} {\bibfnamefont {N.~A.~M.}\ \bibnamefont
  {Ara\'ujo}}, \ and\ \bibinfo {author} {\bibfnamefont {S.~S.}\ \bibnamefont
  {Manna}},\ }\href@noop {} {\bibfield  {journal} {\bibinfo  {journal} {Phys.
  Rev. E}\ }\textbf {\bibinfo {volume} {98}},\ \bibinfo {pages} {062118}
  (\bibinfo {year} {2018})}\BibitemShut {NoStop}%
\bibitem [{\citenamefont {Furlan}\ \emph {et~al.}(2020)\citenamefont {Furlan},
  \citenamefont {dos Santos}, \citenamefont {Ziff},\ and\ \citenamefont
  {Dickman}}]{Furlan2020}%
  \BibitemOpen
  \bibfield  {author} {\bibinfo {author} {\bibfnamefont {A.~P.}\ \bibnamefont
  {Furlan}}, \bibinfo {author} {\bibfnamefont {D.~C.}\ \bibnamefont {dos
  Santos}}, \bibinfo {author} {\bibfnamefont {R.~M.}\ \bibnamefont {Ziff}}, \
  and\ \bibinfo {author} {\bibfnamefont {R.}~\bibnamefont {Dickman}},\
  }\href@noop {} {\bibfield  {journal} {\bibinfo  {journal} {Phys. Rev.
  Research}\ }\textbf {\bibinfo {volume} {2}},\ \bibinfo {pages} {043027}
  (\bibinfo {year} {2020})}\BibitemShut {NoStop}%
\bibitem [{\citenamefont {Pasinetti}\ \emph {et~al.}(2019)\citenamefont
  {Pasinetti}, \citenamefont {Ramirez}, \citenamefont {Centres}, \citenamefont
  {Ramirez-Pastor},\ and\ \citenamefont {Cwilich}}]{Pasinetti2019}%
  \BibitemOpen
  \bibfield  {author} {\bibinfo {author} {\bibfnamefont {P.~M.}\ \bibnamefont
  {Pasinetti}}, \bibinfo {author} {\bibfnamefont {L.~S.}\ \bibnamefont
  {Ramirez}}, \bibinfo {author} {\bibfnamefont {P.~M.}\ \bibnamefont
  {Centres}}, \bibinfo {author} {\bibfnamefont {A.~J.}\ \bibnamefont
  {Ramirez-Pastor}}, \ and\ \bibinfo {author} {\bibfnamefont {G.~A.}\
  \bibnamefont {Cwilich}},\ }\href@noop {} {\bibfield  {journal} {\bibinfo
  {journal} {Physical Review E}\ }\textbf {\bibinfo {volume} {100}},\ \bibinfo
  {pages} {052114} (\bibinfo {year} {2019})}\BibitemShut {NoStop}%
\bibitem [{\citenamefont {Ramirez}\ \emph {et~al.}(2019)\citenamefont
  {Ramirez}, \citenamefont {Centres},\ and\ \citenamefont
  {Ramirez-Pastor}}]{Ramirez2019}%
  \BibitemOpen
  \bibfield  {author} {\bibinfo {author} {\bibfnamefont {L.~S.}\ \bibnamefont
  {Ramirez}}, \bibinfo {author} {\bibfnamefont {P.~M.}\ \bibnamefont
  {Centres}}, \ and\ \bibinfo {author} {\bibfnamefont {A.~J.}\ \bibnamefont
  {Ramirez-Pastor}},\ }\href@noop {} {\bibfield  {journal} {\bibinfo  {journal}
  {Journal of Statistical Mechanics: Theory and Experiment}\ }\textbf {\bibinfo
  {volume} {2019}},\ \bibinfo {pages} {033207} (\bibinfo {year}
  {2019})}\BibitemShut {NoStop}%
\bibitem [{\citenamefont {Cornette}\ \emph {et~al.}(2006)\citenamefont
  {Cornette}, \citenamefont {Ramirez-Pastor},\ and\ \citenamefont
  {Nieto}}]{Cornette2006}%
  \BibitemOpen
  \bibfield  {author} {\bibinfo {author} {\bibfnamefont {V.}~\bibnamefont
  {Cornette}}, \bibinfo {author} {\bibfnamefont {A.}~\bibnamefont
  {Ramirez-Pastor}}, \ and\ \bibinfo {author} {\bibfnamefont {F.}~\bibnamefont
  {Nieto}},\ }\href@noop {} {\bibfield  {journal} {\bibinfo  {journal} {Physics
  Letters A}\ }\textbf {\bibinfo {volume} {353}},\ \bibinfo {pages} {452 }
  (\bibinfo {year} {2006})}\BibitemShut {NoStop}%
\bibitem [{\citenamefont {Kondrat}(2006)}]{Kondrat2006}%
  \BibitemOpen
  \bibfield  {author} {\bibinfo {author} {\bibfnamefont {G.}~\bibnamefont
  {Kondrat}},\ }\href@noop {} {\bibfield  {journal} {\bibinfo  {journal} {The
  Journal of Chemical Physics}\ }\textbf {\bibinfo {volume} {124}},\ \bibinfo
  {pages} {054713} (\bibinfo {year} {2006})}\BibitemShut {NoStop}%
\bibitem [{\citenamefont {Centres}\ and\ \citenamefont
  {Ramirez-Pastor}(2015)}]{Centres2015}%
  \BibitemOpen
  \bibfield  {author} {\bibinfo {author} {\bibfnamefont {P.~M.}\ \bibnamefont
  {Centres}}\ and\ \bibinfo {author} {\bibfnamefont {A.~J.}\ \bibnamefont
  {Ramirez-Pastor}},\ }\href@noop {} {\bibfield  {journal} {\bibinfo  {journal}
  {Journal of Statistical Mechanics: Theory and Experiment}\ }\textbf {\bibinfo
  {volume} {2015}},\ \bibinfo {pages} {P10011} (\bibinfo {year}
  {2015})}\BibitemShut {NoStop}%
\bibitem [{\citenamefont {Tarasevich}\ \emph {et~al.}(2015)\citenamefont
  {Tarasevich}, \citenamefont {Laptev}, \citenamefont {Vygornitskii},\ and\
  \citenamefont {Lebovka}}]{Tarasevich2015}%
  \BibitemOpen
  \bibfield  {author} {\bibinfo {author} {\bibfnamefont {Y.~Y.}\ \bibnamefont
  {Tarasevich}}, \bibinfo {author} {\bibfnamefont {V.~V.}\ \bibnamefont
  {Laptev}}, \bibinfo {author} {\bibfnamefont {N.~V.}\ \bibnamefont
  {Vygornitskii}}, \ and\ \bibinfo {author} {\bibfnamefont {N.~I.}\
  \bibnamefont {Lebovka}},\ }\href@noop {} {\bibfield  {journal} {\bibinfo
  {journal} {Phys. Rev. E}\ }\textbf {\bibinfo {volume} {91}},\ \bibinfo
  {pages} {012109} (\bibinfo {year} {2015})}\BibitemShut {NoStop}%
\bibitem [{\citenamefont {Palacios}\ and\ \citenamefont
  {Gomes}(2020)}]{Palacios2020}%
  \BibitemOpen
  \bibfield  {author} {\bibinfo {author} {\bibfnamefont {G.}~\bibnamefont
  {Palacios}}\ and\ \bibinfo {author} {\bibfnamefont {M.~A.~F.}\ \bibnamefont
  {Gomes}},\ }\href@noop {} {\bibfield  {journal} {\bibinfo  {journal} {Journal
  of Physics A: Mathematical and Theoretical}\ }\textbf {\bibinfo {volume}
  {53}},\ \bibinfo {pages} {375003} (\bibinfo {year} {2020})}\BibitemShut
  {NoStop}%
\bibitem [{\citenamefont {Medina}\ \emph {et~al.}(1989)\citenamefont {Medina},
  \citenamefont {Hwa}, \citenamefont {Kardar},\ and\ \citenamefont
  {Zhang}}]{Medina1989}%
  \BibitemOpen
  \bibfield  {author} {\bibinfo {author} {\bibfnamefont {E.}~\bibnamefont
  {Medina}}, \bibinfo {author} {\bibfnamefont {T.}~\bibnamefont {Hwa}},
  \bibinfo {author} {\bibfnamefont {M.}~\bibnamefont {Kardar}}, \ and\ \bibinfo
  {author} {\bibfnamefont {Y.-C.}\ \bibnamefont {Zhang}},\ }\href@noop {}
  {\bibfield  {journal} {\bibinfo  {journal} {Phys. Rev. A}\ }\textbf {\bibinfo
  {volume} {39}},\ \bibinfo {pages} {3053} (\bibinfo {year}
  {1989})}\BibitemShut {NoStop}%
\bibitem [{\citenamefont {Hewett}\ and\ \citenamefont
  {Behrens}(1990)}]{Hewett1990}%
  \BibitemOpen
  \bibfield  {author} {\bibinfo {author} {\bibfnamefont {T.~A.}\ \bibnamefont
  {Hewett}}\ and\ \bibinfo {author} {\bibfnamefont {R.~A.}\ \bibnamefont
  {Behrens}},\ }\href@noop {} {\bibfield  {journal} {\bibinfo  {journal} {SPE
  Formation Evaluation}\ }\textbf {\bibinfo {volume} {5}},\ \bibinfo {pages}
  {217} (\bibinfo {year} {1990})}\BibitemShut {NoStop}%
\bibitem [{\citenamefont {Lauritsen}\ \emph {et~al.}(1993)\citenamefont
  {Lauritsen}, \citenamefont {Sahimi},\ and\ \citenamefont
  {Herrmann}}]{Lauritsen1993}%
  \BibitemOpen
  \bibfield  {author} {\bibinfo {author} {\bibfnamefont {K.~B.}\ \bibnamefont
  {Lauritsen}}, \bibinfo {author} {\bibfnamefont {M.}~\bibnamefont {Sahimi}}, \
  and\ \bibinfo {author} {\bibfnamefont {H.~J.}\ \bibnamefont {Herrmann}},\
  }\href@noop {} {\bibfield  {journal} {\bibinfo  {journal} {Phys. Rev. E}\
  }\textbf {\bibinfo {volume} {48}},\ \bibinfo {pages} {1272} (\bibinfo {year}
  {1993})}\BibitemShut {NoStop}%
\bibitem [{\citenamefont {Schrenk}\ \emph {et~al.}(2013)\citenamefont
  {Schrenk}, \citenamefont {Pos\'e}, \citenamefont {Kranz}, \citenamefont {van
  Kessenich}, \citenamefont {Ara\'ujo},\ and\ \citenamefont
  {Herrmann}}]{Schrenk2013}%
  \BibitemOpen
  \bibfield  {author} {\bibinfo {author} {\bibfnamefont {K.~J.}\ \bibnamefont
  {Schrenk}}, \bibinfo {author} {\bibfnamefont {N.}~\bibnamefont {Pos\'e}},
  \bibinfo {author} {\bibfnamefont {J.~J.}\ \bibnamefont {Kranz}}, \bibinfo
  {author} {\bibfnamefont {L.~V.~M.}\ \bibnamefont {van Kessenich}}, \bibinfo
  {author} {\bibfnamefont {N.~A.~M.}\ \bibnamefont {Ara\'ujo}}, \ and\ \bibinfo
  {author} {\bibfnamefont {H.~J.}\ \bibnamefont {Herrmann}},\ }\href@noop {}
  {\bibfield  {journal} {\bibinfo  {journal} {Phys. Rev. E}\ }\textbf {\bibinfo
  {volume} {88}},\ \bibinfo {pages} {052102} (\bibinfo {year}
  {2013})}\BibitemShut {NoStop}%
\bibitem [{\citenamefont {Makse}\ \emph {et~al.}(1996)\citenamefont {Makse},
  \citenamefont {Havlin}, \citenamefont {Schwartz},\ and\ \citenamefont
  {Stanley}}]{Makse1996}%
  \BibitemOpen
  \bibfield  {author} {\bibinfo {author} {\bibfnamefont {H.~A.}\ \bibnamefont
  {Makse}}, \bibinfo {author} {\bibfnamefont {S.}~\bibnamefont {Havlin}},
  \bibinfo {author} {\bibfnamefont {M.}~\bibnamefont {Schwartz}}, \ and\
  \bibinfo {author} {\bibfnamefont {H.~E.}\ \bibnamefont {Stanley}},\
  }\href@noop {} {\bibfield  {journal} {\bibinfo  {journal} {Phys. Rev. E}\
  }\textbf {\bibinfo {volume} {53}},\ \bibinfo {pages} {5445} (\bibinfo {year}
  {1996})}\BibitemShut {NoStop}%
\bibitem [{\citenamefont {Prakash}\ \emph {et~al.}(1992)\citenamefont
  {Prakash}, \citenamefont {Havlin}, \citenamefont {Schwartz},\ and\
  \citenamefont {Stanley}}]{Prakash1992}%
  \BibitemOpen
  \bibfield  {author} {\bibinfo {author} {\bibfnamefont {S.}~\bibnamefont
  {Prakash}}, \bibinfo {author} {\bibfnamefont {S.}~\bibnamefont {Havlin}},
  \bibinfo {author} {\bibfnamefont {M.}~\bibnamefont {Schwartz}}, \ and\
  \bibinfo {author} {\bibfnamefont {H.~E.}\ \bibnamefont {Stanley}},\
  }\href@noop {} {\bibfield  {journal} {\bibinfo  {journal} {Phys. Rev. A}\
  }\textbf {\bibinfo {volume} {46}},\ \bibinfo {pages} {R1724} (\bibinfo {year}
  {1992})}\BibitemShut {NoStop}%
\bibitem [{\citenamefont {Zierenberg}\ \emph {et~al.}(2017)\citenamefont
  {Zierenberg}, \citenamefont {Fricke}, \citenamefont {Marenz}, \citenamefont
  {Spitzner}, \citenamefont {Blavatska},\ and\ \citenamefont
  {Janke}}]{Zierenberg2017}%
  \BibitemOpen
  \bibfield  {author} {\bibinfo {author} {\bibfnamefont {J.}~\bibnamefont
  {Zierenberg}}, \bibinfo {author} {\bibfnamefont {N.}~\bibnamefont {Fricke}},
  \bibinfo {author} {\bibfnamefont {M.}~\bibnamefont {Marenz}}, \bibinfo
  {author} {\bibfnamefont {F.~P.}\ \bibnamefont {Spitzner}}, \bibinfo {author}
  {\bibfnamefont {V.}~\bibnamefont {Blavatska}}, \ and\ \bibinfo {author}
  {\bibfnamefont {W.}~\bibnamefont {Janke}},\ }\href@noop {} {\bibfield
  {journal} {\bibinfo  {journal} {Phys. Rev. E}\ }\textbf {\bibinfo {volume}
  {96}},\ \bibinfo {pages} {062125} (\bibinfo {year} {2017})}\BibitemShut
  {NoStop}%
\bibitem [{\citenamefont {Schrenk}\ \emph {et~al.}(2012)\citenamefont
  {Schrenk}, \citenamefont {Ara{\'{u}}jo}, \citenamefont {{Andrade Jr}},\ and\
  \citenamefont {Herrmann}}]{Schrenk2012}%
  \BibitemOpen
  \bibfield  {author} {\bibinfo {author} {\bibfnamefont {K.~J.}\ \bibnamefont
  {Schrenk}}, \bibinfo {author} {\bibfnamefont {N.~A.~M.}\ \bibnamefont
  {Ara{\'{u}}jo}}, \bibinfo {author} {\bibfnamefont {J.~S.}\ \bibnamefont
  {{Andrade Jr}}}, \ and\ \bibinfo {author} {\bibfnamefont {H.~J.}\
  \bibnamefont {Herrmann}},\ }\href@noop {} {\bibfield  {journal} {\bibinfo
  {journal} {Scientific Reports}\ }\textbf {\bibinfo {volume} {2}},\ \bibinfo
  {pages} {348} (\bibinfo {year} {2012})}\BibitemShut {NoStop}%
\bibitem [{\citenamefont {Cardy}(2006)}]{Cardy2006}%
  \BibitemOpen
  \bibfield  {author} {\bibinfo {author} {\bibfnamefont {J.}~\bibnamefont
  {Cardy}},\ }\href@noop {} {\bibfield  {journal} {\bibinfo  {journal} {Journal
  of Statistical Physics}\ }\textbf {\bibinfo {volume} {125}},\ \bibinfo
  {pages} {1} (\bibinfo {year} {2006})}\BibitemShut {NoStop}%
\bibitem [{\citenamefont {Ziff}(2011)}]{Ziff2011}%
  \BibitemOpen
  \bibfield  {author} {\bibinfo {author} {\bibfnamefont {R.~M.}\ \bibnamefont
  {Ziff}},\ }\href@noop {} {\bibfield  {journal} {\bibinfo  {journal} {Phys.
  Rev. E}\ }\textbf {\bibinfo {volume} {83}},\ \bibinfo {pages} {020107(R)}
  (\bibinfo {year} {2011})}\BibitemShut {NoStop}%
\end{thebibliography}

%
\end{document}